\documentstyle[12pt,aaspp4,psfig,flushrt,tighten]{article}

\lefthead{Schneider, Ferrara, Natarajan \& Omukai}
\righthead{First Stars, Very Massive Black Holes and Cosmic Metals}

\def\be{\begin{equation}}
\def\ee{\end{equation}}

\def\msun{{M_\odot}}

\def\etal{{\it et al.~}}

\def\HH{${\rm {H_2}}\,\,$}

\def\sngg{SN$_{\gamma\gamma}$~}
\def\fgg{f_{\gamma\gamma}}

\def\cm{${\rm cm}$}

\def\gsim{\lower.5ex\hbox{\gtsima}}
\def\lsim{\lower.5ex\hbox{\ltsima}}
\def\gtsima{$\; \buildrel > \over \sim \;$}
\def\ltsima{$\; \buildrel < \over \sim \;$}
\def\prosima{$\; \buildrel \propto \over \sim \;$}
\def\gsim{\lower.5ex\hbox{\gtsima}}
\def\lsim{\lower.5ex\hbox{\ltsima}}
\def\simgt{\lower.5ex\hbox{\gtsima}}
\def\simlt{\lower.5ex\hbox{\ltsima}}
\def\simpr{\lower.5ex\hbox{\prosima}}

\def\ie{{\frenchspacing\it i.e. }}
\def\eg{{\frenchspacing\it e.g. }}

\def\beq#1{\begin{equation}\label{#1}}
\def\eeq{\end{equation}}
\def\beqa#1{\begin{eqnarray}\label{#1}}
\def\eeqa{\end{eqnarray}}

\def\Ls{\ L_{\odot}}
\def\K{{\rm \ K}}

\def\cm{{\rm \ cm}}

\def\HH{H$_2$ }
\def\H2p{H$_2^+$ }

\def\mH2p{H_2^+}

\input psfig
\begin{document}

\title{First stars, very massive black holes and metals}

\author{R. Schneider$^1$, A. Ferrara$^1$, P. Natarajan$^{2,3}$, 
and K. Omukai$^{1,4}$}

\affil{\footnotesize $^1$ Osservatorio Astrofisico di Arcetri, Largo Enrico Fermi 5, 
	  50125 Firenze, Italy}
\affil{\footnotesize $^2$ Department of Astronomy, Yale University, New Haven, 
CT 06520-8101, USA}
\affil{\footnotesize $^3$ Yale Center for Astronomy and Astrophysics, Yale University, New Haven, CT 06511, USA}
\affil{\footnotesize $^4$ Division of Theoretical Astrophysics, National Astronomical 
Observatory, Mitaka, Tokyo 181-8588, Japan} 

\begin{abstract}
Recent studies suggest that the initial mass function (IMF) of the
first stars (PopIII) was likely to be extremely top-heavy, unlike what is
observed at present. We propose a scenario to generate fragmentation
to lower masses once the first massive stars have formed and derive
constraints on the primordial IMF. We estimate the mass fraction of
pair-unstable supernovae (SN$_{\gamma\gamma}$), shown to be the dominant sources of the
first heavy elements. These metals enrich the surrounding gas up to $\approx
10^{-4}\,Z_{\odot}$, when a transition to efficient cooling-driven
fragmentation producing $ \simlt 1 \msun$ clumps occurs.  We argue
that the remaining fraction of the first stars ends up in $\approx 100
\msun$ VMBHs (Very Massive Black Holes). If we further assume that all
these VMBHs are likely to end up in the centers of galactic nuclei
constituting the observed SMBHs, then $\approx 6\%$ of the first stars
contributed to the initial metal enrichment and the IMF remained
top-heavy down to a redshift $z \approx$ 18.
5. At the other
extreme if none of these VMBHs have as yet ended up in SMBHs, we
expect them to be either (i) en-route towards galactic nuclei thereby
accounting for the X-ray bright off-center sources detected locally by
ROSAT or (ii) as the dark matter candidate composing the entire
baryonic halos of galaxies. For case (i) we expect all but a negligible 
fraction of the primordial stars to produce metals causing the transition at the
maximum possible redshift of $\sim 22.1$, and for case (ii) $\sim 3 \times 10^{-5}$ - a very
negligible fraction of the initial stars produce the metals and the
transition redshift occurs at $z_f \sim 5.4$. In this paper we present
a framework (albeit one that is not stringently constrained at
present) that relates the first episode of star formation to the fate
of their remnants at late times. Clearly, further progress in
understanding the formation and fragmentation of PopIII stars
within the cosmological context will provide tighter constraints in
the future. We conclude with a discussion of several hitherto
unexplored implications of a high-mass dominated star formation mode
in the early Universe.
\end{abstract}


\section{Introduction}

Recent studies have started to tackle the formation and collapse of
the first cosmic structures (often referred to as PopIII objects)
through numerical simulations (Abel \etal 1998; Bromm, Coppi \& Larson
1999; Abel, Bryan \& Norman 2000; Bromm \etal 2001) based on
hierarchical scenarios of structure formation. These studies have
shown that gravitational collapse induces fragmentation of pre-galactic
units with an initial baryonic mass $\approx 10^5 \msun$ into smaller
clumps with a typical mass of $10^{3} \msun$, which corresponds to the
Jeans mass set by molecular hydrogen cooling. However, a considerable
mass range ($10^{2-4} M_\odot$) for the clumps seems plausible.

Tracking the subsequent gravitational collapse of these metal free
clumps is a very challenging problem as it requires the simultaneous
solution of hydrodynamics and of (cooling lines) radiative transfer
(Omukai \& Nishi 1998; Nakamura \& Umemura 1999; Ripamonti \etal
2001). Preliminary attempts and several physical arguments indicate
that these clumps do not fragment into smaller units as the evolution
progresses to higher densities.  Independent studies (Hernandez \&
Ferrara 2001) comparing the observed number of metal poor stars with
that predicted by cosmological models, also imply that the
characteristic stellar mass sharply increases with redshift. Hence,
there are grounds to believe that the first stars were very massive.

The evolution of massive, metal free stars is currently subject to a
rapidly growing number of studies (Fryer 1999; Fryer, Woosley \& Heger
2001; Heger \& Woosley 2001) which rejuvenate earlier activity (Fowler
\& Hoyle 1964; Carr, Bond \& Arnett 1984; El Eid \etal 1983; Fricke
1973; Fuller \etal 1986). As we discuss in detail later in the paper, 
stars more massive than about 260 $M_\odot$ 
collapse completely to black holes, therefore not contributing to the
metal enrichment of the surrounding gas. Similar arguments apply to
stars in a lower mass window (30 $M_\odot$ - 140 $M_\odot$), which are
also expected to end their evolution as black holes. Hence, if
supernovae from more standard progenitors (stellar masses in the range
8 $M_\odot$ - 40 $M_\odot$) are either not formed efficiently or occur
in negligible numbers, it appears that the initial cosmic metal
enrichment had to rely on the heavy element yield from the so-called
pair-unstable supernovae (\sngg), whose explosion leaves no
remnant. This conclusion can potentially be tested by studying
peculiar elemental abundances, for example of heavy {\it r-}process
elements (Oh \etal 2001).

Metallicity is thought to noticeably affect the fragmentation
properties of a gravitationally unstable gas. For example, Bromm \etal
(2001) have shown that the evolution of a collapsing proto-galaxy depends
strongly on the level of gas pre-enrichment. These authors argue that
a critical metallicity $\approx 10^{-4} Z_\odot$ exists such that,
above that value vigorous fragmentation into relatively low mass
($\approx 10 M_\odot$) clumps takes place, differently from what is
discussed above under metal-free conditions.

The key question then concerns the interplay between the properties of
the IMF and the metal enrichment of the gas. To better illustrate
this, let us ideally suppose that the first stars are all formed -- as
suggested by numerical simulations -- with masses above the \sngg mass
threshold, thus leading to the formation of black holes.  Then,
because metals are completely swallowed by the latter, the gas retains
its primordial composition and star formation continues in the
high-mass biased mode. A solution to this "star formation conundrum"
must evidently exist as at present stars form with a much lower
characteristic mass ($\approx 1 M_\odot$).

In this paper, we describe in detail a possible solution to this
conundrum utilizing various constraints: metal abundance patterns in
clusters, the mass density of super-massive black holes, off-nuclear
galactic X-ray sources and, to a more speculative extent, the nature
of a particular class (optically hidden) of gamma-ray bursts - we
attempt to infer the main properties of the early stages of cosmic
star formation.

\section{Fragmentation modes}

In this Section, we focus on the first episode of star formation and
discuss the relevant cooling criteria and timescales in detail. 
We work within the paradigm of hierarchical cold
dark matter (CDM) models for structure formation, wherein dark matter
halos collapse and the baryons in them condense, cool, and eventually
form stars. Once a halo of mass $M$ collapses at redshift of
$z_{\rm c}$, the baryons are shock-heated to the virial
temperature given by
\begin{equation}
T_{\rm vir} = 10^{4.5} \, \mu \, M_8^{2/3} 
\left({1+z_{\rm c}\over 10}\right) \K,
\end{equation}
where $M_8=M/10^8 h^{-1} \msun$ and $\mu$ is the molecular weight.  If
$T_{\rm vir} \gsim 10^4 \K$ [or equivalently $M \gsim 10^9 (1+z_{\rm
c})^{-3/2} h^{-1} \msun$], the baryonic gas cools due to the
excitation of hydrogen Ly$\alpha$ line. In the absence of metals, as
expected for the very first episode of star formation, objects with
$T_{\rm vir} \lsim 10^4 \K$ can cool only through the collisional
excitation of molecular hydrogen. Hereafter, in this work, we refer to
the former objects as Ly$\alpha$-cooling halos and to the latter as
PopIII objects.

The first stars form once the gas cools.  The typical initial mass
function (IMF) of this first generation of stars is still highly
uncertain. Several authors have tackled this crucial issue through
theoretical (Rees 1976; Rees \& Ostriker 1977; Silk 1977, 1983;
Haiman, Thoul \& Loeb 1996; Uehara \etal 1996) and numerical
approaches (Omukai \& Nishi 1998; Nakamura \& Umemura 1999, 2001;
Abel, Bryan \& Norman 2000; Bromm, Coppi \& Larson 2001; Omukai 2000,
2001; Ripamonti \etal 2001).  Recent multi-dimensional simulations of
the collapse and fragmentation of primordial gas within PopIIIs, show preliminary
indications that the IMF could be either top-heavy with typical masses
of order $\gsim 10^2 \msun$ (Abel, Bryan \& Norman 2000; Bromm, Coppi
\& Larson 2001) or bimodal with peaks at $\approx 10^2 \msun$
and $\approx 1 \msun$ (Nakamura \& Umemura 2001). Note, that these numerical
treatments cannot address the issue of the mass spectrum.  Therefore,
it is important to understand what physical processes set the scales
of fragment masses and hence the stellar mass spectrum.
 
It is obvious that much hinges on the physics of cooling, primarily
the number of channels available for the gas to cool and the
efficiency of the process (Rees \& Ostriker 1977). In general, cooling
is efficient when the cooling time  $t_{\rm
cool} = 3 n k T/2 \Lambda(n,T)$ is much shorter than the
free-fall time $t_{\rm ff} = (3 \pi/32 G \rho)^{1/2}$, i.e.
$t_{\rm cool} \ll t_{\rm ff}$; where $n$ ($\rho$) is the 
gas number (mass) 
density and $\Lambda(n,T)$ is the net radiative cooling rate 
[erg $\mbox{cm}^{-3} \mbox{s}^{-1}$].  This efficiency condition
implies that the energy deposited by gravitational contraction cannot
balance the radiative losses; as a consequence, temperature decreases with
increasing density. Under such circumstances, the cloud cools and then
fragments. At any given time, fragments form on a scale that is small
enough to ensure pressure equilibrium at the corresponding
temperature, \ie the Jeans length scale, \be R_F \approx \lambda_J
\propto c_{\rm s} t_{\rm ff} \propto n^{\gamma/2 -1}
\label{eq:Rf}
\ee 
where the sound speed $c_{\rm s} = (RT/\mu)^{1/2}$, $T \propto
n^{\gamma-1}$, where $\gamma$ is the adiabatic index. 
Since $c_{\rm s}$ varies on the cooling timescale,
the corresponding $R_F$ becomes smaller as $T$
decreases. Similarly, the corresponding fragment mass is the Jeans
mass, 
\be M_F \propto n R_F^{\eta} \propto n^{\eta\gamma/2 
+ (1-\eta)},
\label{eq:Jm}
\ee with $\eta = 2$ for filaments and $\eta = 3$ for spherical
fragments (Spitzer 1978). This hierarchical fragmentation process
comes to an end when  cooling becomes
inefficient because {\it (i)} the critical density for LTE is reached 
or {\it (ii)} the gas becomes optically thick to 
cooling radiation; in both cases, at that juncture 
$t_{\rm cool} \gsim t_{\rm ff}$. At this stage, the
temperature cannot decrease any further and it either remains constant
(if energy deposition by gravitational contraction is exactly balanced
by radiative losses) or increases. The necessary condition to stop
fragmentation and start gravitational contraction within each fragment
is that the Jeans mass does not decrease any further, thus favoring
fragmentation into sub-clumps. From eq.~\ref{eq:Jm}, this implies 
the following condition:
\be 
\gamma \gsim 2 \frac{\eta-1}{\eta}, 
\label{cond}
\ee
which translates into $\gamma \gsim 4/3$ for a spherical fragment and
$\gamma \gsim 1$ for a filament. Thus, a filament is marginally stable
and contracts quasi-statically when, $t_{\rm cool} \approx
t_{\rm ff}$, and the gas becomes isothermal. Finally,
when $t_{\rm cool}\gg t_{\rm ff}$ or the fragments become optically
thick to cooling radiation, the temperature increases as the
contraction proceeds adiabatically.
 
\subsection{Fragmentation of metal-free clouds}

In this subsection, we follow the evolution of metal-free primordial
clumps during the fragmentation process. These results are based on
the model of Omukai (2001).  The gas within the dark matter halo is
given an initial temperature of $100 \K$ and the subsequent thermal
and chemical evolution of the gravitationally collapsing cloud is
followed numerically until a central proto-stellar core forms.

The gas within the dark matter halo gets shock-heated to the virial
temperature $T_{\rm vir}$, which is typically $\gg 100 \K$.  However,
after a short transient phase (irrelevant for the present analysis),
the evolutionary track in the $(n,T)$ plane shown in Fig.~1 (top
curve, upper panel) provides a good description of the thermal
evolution of the gas.  The metal-free gas is able to cool down to
temperatures of a few hundred Kelvin regardless of the initial virial
temperature. This is the minimum temperature at which molecular line
cooling becomes effective.

In this analysis, external sources of heating are not included (\ie external
UV background or CMB radiation). This is because at redshifts prior 
to reionization the UV background
field is relatively weak and inhomogeneous (Ciardi, Ferrara, \& Abel
2000), hence we do not expect it to be important.  Also, the
CMB energy input is negligible for clouds with $T > 100$~K (or $Z <
10^{-4} Z_\odot$) where the first stars presumably formed at redshifts
$z <30$. It might affect the temperature evolution of more metal rich systems
shown in Fig.~1, but does not modify the conclusions drawn here. 

The gas within halos with $T_{\rm vir} > 10^4$~K starts to cool
through the hydrogen Ly$\alpha$ line. It quickly reaches a temperature
of $\approx 8000\,\K$ and, at this stage, the fraction of molecular
hydrogen formed is sufficient to activate H$_2$ ro-vibrational line
cooling.  Thereafter, the gas within a Ly$\alpha$-cooling halo follows
the same thermal evolution as the gas within PopIIIs. Independent of
the virial temperature, the thermal evolution of the gas rapidly
converges to the $(n,T)$ track corresponding to $Z=0$ (the zero
metallicity track), shown in Fig.~1.  The temperature of the gas
decreases with increasing density, thus favoring fragmentation into
sub-clumps.

As the number density increases, it reaches the critical value $n_{\rm
cr} \approx 10^3 \,\mbox{cm}^{-3}$; and the corresponding Jeans mass
is $ \approx 10^4 \msun$ (empty dot in the upper panel of Fig.~1). The
cooling time at this critical point becomes comparable to the free
fall time as a consequence of the \HH levels being populated according
to LTE [regime where the cooling rate $\Lambda(n,T) \propto n$] and no
longer according to NLTE [regime where $\Lambda(n,T)\propto n^2$].
The temperature then starts to rise slowly.

At this stage, the stability of the fragments toward further increase
in the density needs to be investigated according to the condition
eq. \ref{cond} above. The lower panel of Fig.~1 shows the density
dependence of $\gamma$ (see eq.~\ref{eq:Rf}) for each metallicity
track.  For a metal-free gas $\gamma$ lies in the range
$0<\gamma-1<1/3$, implying that further fragmentation is unlikely to
occur unless the fragments are spherical. Although the gravitational
evolution will probably favor a tendency towards spherical symmetry,
this is not likely to occur until the central density has reached high
($\approx 10^8~{\rm cm}^{-3}$) values as seen from simulations (\eg Abel \etal 2000). 

The only two deviations
from the above range occur (see lower panel of Fig.~1) around $n =
10^{10}{\rm cm}^{-3}$ and $10^{16}{\rm cm}^{-3}$.  The former is a
result of the thermal instability due to three-body H$_{2}$ formation
(Silk 1983; Haiman et al. 1996).  However, this instability is quite
weak and does not lead to fragmentation (Abel et al. 2000).  The
latter (at $n = 10^{16}{\rm cm}^{-3}$) is caused by H$_{2}$
collision-induced emission. This instability is also weak and probably
unimportant. 

It is important to stress that, even if the fragments are nearly
spherical, fragmentation will be modest and is likely to result only
in a low multiplicity stellar system.  As the density increases,
quasi-static contraction takes place ($n = 10^{20}{\rm cm}^{-3}$)
until the fragments become optically thick to \HH lines, $t_{\rm cool}
\gg t_{\rm ff}$, and adiabatically collapse to increasingly higher
central densities and temperatures. At this stage, $\gamma > 4/3$ and
a central hydro-static core (stellar core, filled circle in Fig.~1,
upper panel) is formed, with mass $\approx\,10^{-3} \msun$ (Omukai \&
Nishi 1998).

We stress again that, as long as the gas is metal-free, these sequence
of events and conclusions hold independent of the halo virial
temperature, \ie for PopIIIs as well as for Ly$\alpha$-cooling halos.
In Fig.~2 we show the evolution with temperature of the ionization
fraction and of the fraction of molecular hydrogen for typical
Ly$\alpha$-cooling halos of mass $M = 10^8 \msun$ at three different
redshifts, $z = 15, 20, 25$. We find that the evolution of these
fractions is independent of the mass (or, equivalently, $T_{\rm vir}$)
and virialization redshift.  Molecule formation in the post-shock flow
which follows the virialization of the gas within a dark matter halo
has been recently investigated by Uehara \& Inutsuka (2000). If the
gas is fully ionized, molecules form through non-equilibrium
recombination, leading to overall fractions that are much higher than
in the expanding homogeneous universe. As a consequence of enhanced
\HH and HD fractions, the gas rapidly cools to well below $10^4 \K$ and
fragments. Molecular chemistry is important also for star
formation in halos with $T_{vir} > 10^4$~K as assessed 
by Susa \etal (1998) and Oh \& Haiman (2001). They find that initial atomic line cooling leaves
a large residual free electron density which allows molecule formation up to
a universal fraction of $x_{\rm{H}_2} = 10^{-3}$. The newly formed molecules cool
the gas further to $\sim 100$~K and the gas fragments on mass scales of a few $100 \msun$.

Once the critical density for \HH ($n_{\rm cr} \approx 10^3
\mbox{cm}^{-3}$ or $10^5 \mbox{cm}^{-3}$ for HD, if this is assumed to
be the main coolant) has been reached, LTE conditions for the level
populations disfavor further cooling and fragmentation. At this stage,
fragments have masses comparable to the Jeans mass corresponding to
the point $(n_{\rm cr}, T_{\rm cr})$ and virialize, following the
evolution described above. Therefore, independent of the initial
virial temperature, fragments are formed with typical masses $\approx
10^3 - 10^4 \msun$.

Each fragment is characterized by a central core of $10^{-3} \msun$
surrounded by a large envelope of gravitationally unstable gas. The
core grows in mass due to gas accretion from the envelope.  The mass
of the formed stars depends on the accretion rate as well as on the
fragment mass (Larson 1999). In the absence of metals, radiation
pressure cannot counter-balance mass accretion onto the core
(Ripamonti \etal 2001, Omukai \& Inutsuka 2001) and the mass of the resulting star is
comparable to that of the original fragment, \ie $\lsim 10^3
\msun$. We will return to this crucial issue in Section
\ref{sec:accr}.  Of course, if the parent cloud is rotating, angular momentum,
if not dissipated, might prevent the collapse of the entire cloud. However, for
the arguments relevant to the present paper, it is sufficient that the overall efficiency
is of a few \% thus yielding a few hundreds solar mass stars out of the $\approx 10^4 \msun$ 
fragments predicted by Fig.~1. 
The question that needs to be addressed now is what
mechanism can finally lead to the transition to a conventional mode of
star formation, \ie to a standard IMF, and what are the necessary
conditions for this to occur.

\subsection{Fragmentation of metal-enriched clouds}

We now consider the effects of the presence of heavy elements on the
fragmentation process.  The discussion presented in this Section is based
on the analysis by Omukai (2000) (see the original paper for details). 
Fig.~1, reproduced from Omukai (2000), shows the effects of metal enrichment
on the $(n,T)$ tracks for the same initial conditions and different
values of the mean metallicity.

In general, clouds with lower metallicity tend to be warmer because of
their lower radiative cooling ability.  As long as the clouds are
transparent, cooling and fragmentation occur. Clouds with a mean
metallicity $Z \approx 10^{-6} Z_\odot$ follow the same evolution as
that of the gas with primordial composition in the $(n,T)$
plane. However, at $Z \gsim 10^{-4}Z_\odot$, \HH formation on grain
surfaces enhances cooling at low density. Dust grains are well known to
condense out the ejecta of SNe (see Todini \& Ferrara 2001 and references therein).  
When the LTE-NLTE transition
occurs for \HH, the cloud can still cool (though less efficiently) due
to OI line cooling.  At densities $> 10^{6} \cm^{-3}$, heating due to
\HH formation becomes larger than compressional work, i.e. $\gamma >
1$ and the temperature starts to increase until thermal emission from
grains due to energy transfer between gas and dust dominates the
cooling. This occurs at a density $n \approx 10^{10} \cm^{-3}$, where
the temperature drops to $\approx 100 \K$ and a new fragmentation
phase occurs. This shows up in the lower panel of Fig.~1 as the large
dip in the $\gamma$ evolution. The minimum fragment mass is reached at
the point indicated by the empty circle, when the density is $\approx
10^{13}\,{\rm cm^{-3}}$ and the corresponding Jeans mass is of order
$10^{-2} \msun$.  Finally, as the density increases, the gas becomes
opaque to dust thermal emission, fragmentation stops and compressional
heating causes the fragments to contract adiabatically ($\gamma >
4/3$).  Therefore, a critical metallicity of $ Z_{\rm cr} \approx
10^{-4} Z_{\odot}$ can be identified, which marks the transition point
between metal-free and metal-rich gas evolution.

When the metallicity is $Z>10^{-4}Z_\odot$, at density $\lsim 10^4
\cm^{-3}$ cooling is driven by OI, CI and CO line emission. When the
NLTE-LTE transition for the level populations of CO occurs,
fragmentation stops and the temperature increases due to \HH
formation. The larger concentration of dust grains (assumed here to be
proportional to the mean metallicity) leads to a significant thermal
emission which is responsible for cooling the gas and starting a new
phase of fragmentation. This stops when $T_{grain} \simeq T$ and
thereafter the fragments contract quasi-statically until they become
optically opaque to dust emission and adiabatic contraction
occurs. Due to the enhanced ability to cool, fragmentation stops at
lower temperatures and densities for higher metallicity clouds (see
open circles in the lower panel of Fig.~1). However, the Jeans mass
corresponding to the minimum fragmentation scale is always $10^{-2}
\msun \le M_F \le 1 \msun$ for the metallicity range $10^{-4} \le
Z/Z_{\odot} \le 1$, several orders of magnitude smaller than for a
cloud with no metals.
 
At the onset of adiabatic contraction, when $\gamma$ becomes larger
than 4/3, an initial hydro-static core forms (transient core) with a
mass $\approx 10^{-2} \msun$ regardless of the metallicity of the
gas. This transient core is fully molecular and is absent when the gas
is metal-free. The temperature of the transient core increases as its
mass increases due to accretion of surrounding gas.  Eventually, the
temperature reaches about $2000 \K$, where H$_{2}$ dissociation
begins.  This softens the equation of state until the dissociation is
almost complete. Then, $\gamma$ falls below 4/3 in the density range
$10^{16}-10^{20}\,{\rm cm^{-3}}$.  Note that the thermal evolution
after H$_{2}$ dissociation (i.e., $n >10^{16}\,{\rm cm^{-3}}$) is the
same independent of the initial composition of the gas.  After
$\gamma$ once again exceeds 4/3, a hydro-static core (so-called
stellar core) forms.  Its physical characteristics are independent of
metallicity ($n \approx\,10^{22}\,{\rm cm^{-3}}, M \approx\,10^{-3}
\msun$; Omukai 2000).

\subsection{Formation of proto-stars}
\label{sec:accr}

Proto-stars, whose mass is initially very low (about $10^{-3} \msun$),
grow in mass by accretion of the envelope material.  The final mass of
stars is, then, determined not only by the mass of the fragments, but
also by when accretion stops. The accretion rate onto the proto-star is
related to the sound speed (and therefore, temperature) of the 
proto-stellar cloud by the relation, $\dot{M} \simeq c_{\rm s}^{3}/G$, 
(Stahler, Shu, \& Taam 1980).

As seen in Fig.~1, the temperature of proto-stellar clouds decreases
with metallicity. Therefore, the mass accretion rate is higher for
proto-stars formed from lower metallicity gas.  If dust is present in
the accretion flow, accretion onto the massive proto-star becomes
increasingly difficult owing to the radiation pressure onto the
dust. In present-day interstellar gas, accretion onto stars more
massive than $30 \msun$ is inhibited by this mechanism (Wolfire \&
Cassinelli 1987). In gas with lower metallicity, the mass bound is
expected to be higher because of the higher accretion rate and lower
radiation force.  In particular, for metal-free gas, this mechanism
does not work.  Therefore, without dust, accretion is likely to
continue until the ambient gas supply is exhausted (Ripamonti \etal
2001, Omukai \& Inutsuka 2001).

In conclusion, the presence of metals not only enables fragmentation
down to smaller mass scales $10^{-2} \msun$, but also breaks the
one-to-one correspondence between the mass of the formed star and that
of the parent fragment by halting the accretion through radiation
force onto the dust.

\section{The star formation conundrum}

According to the scenario proposed above, the first stars that form
within PopIII and Ly$\alpha$-cooling halos out of gas of primordial
composition tend to be very massive, with masses $\approx 10^{2-3}
\msun$.  It is only when metals change the composition of the gas that
further fragmentation occurs producing stars with significantly lower
masses. It is at this stage that we expect a transition to occur from
a top-heavy IMF towards a more conventional IMF with a wide range of
masses, such as the one observed locally (Scalo 1998, Kroupa 2001).

A growing body of observational evidence points to an early top-heavy
IMF (see Hernandez \& Ferrara 2000 and references therein).
Compelling arguments for an early top-heavy IMF can also be made from
observations on various scales: {\it (i)} the early enrichment of our
Galaxy required to solve the so-called G-dwarf problem, {\it (ii)} the
abundance patterns of metals in the Intracluster Medium (ICM), {\it
(iii)} the energetics of the ICM, {\it (iv)} the non-detection of
metal-free stars, and {\it (v)} the over-production of low mass stars
at the present epoch and metals at $z \sim 2 - 5$ for the submillimeter
derived star formation histories using SCUBA detections for a standard
IMF can be resolved with an early top-heavy IMF.

For instance, the ICM metal abundances measured from Chandra and XMM
spectral data are higher than expected from the enrichment by standard
IMF SN yields in cluster galaxy members, which can be explained by a
top-heavy early IMF.  Furthermore, the observed abundance anomalies
(\eg oxygen) in the ICM can be explained by an early generation of
PopIII SNe (Loewenstein 2001). There is observational evidence from
the abundance ratio patterns of [Si/Fe], [Mg/Fe], [Ca/Fe] and [Ti/Fe]
in the extremely metal-poor double-lined spectroscopic binary CS
22876-032 in the halo of our galaxy (Norris, Beers \& Ryan 2000) for
enrichment by a massive, zero-metallicity supernova on comparison with
the theoretical models of Woosley \& Weaver (1995).  These issues,
highly suggestive of top-heavy early star formation, have recently
motivated a series of numerical investigations of the nucleosynthesis
and final fate of metal-free massive stars (Heger \& Woosley 2001;
Fryer, Woosley \& Heger 2001; Umeda \& Nomoto 2002). In their recent paper, Heger \& Woosley
delineate three mass ranges characterized by distinct evolutionary
paths:

\begin{enumerate}

\item $M_\star \gsim 260 \msun$: the nuclear energy release from the
collapse of stars in this mass range is insufficient to reverse the implosion.
The final result is a very massive black hole (VMBH) locking up
all heavy elements produced.

\item $140 \msun \lsim M_\star \lsim 260 \msun$: the mass regime of the 
pair-unstable supernovae (\sngg). Pre-collapse winds and pulsations result 
in little mass loss, the star implodes to a maximum temperature that 
depends on its mass and then explodes, leaving no remnant. The explosion
expels metals into the surrounding ambient ISM.

\item $30 \msun \lsim M_\star \lsim 140 \msun$: black hole formation is
the most likely outcome, because either a successful outgoing shock 
fails to occur or the shock is so weak that the fall-back converts the 
neutron star remnant into a black hole (Fryer 1999). 

\end{enumerate} 

Stars that form in the mass ranges (1) and (3) above fail to eject
most of their heavy elements. If the first stars have masses in excess
of $ 260 \msun$ (in agreement with numerical findings), they
invariably end their lives as VMBHs (in the following we will refer to
VMBHs as black holes of hundred solar mass or so) and do not release
any of their synthesized heavy elements.  However - as we have shown -
as long as the gas remains metal-free, the subsequent generations of
stars will continue to be top-heavy.  This {\it star formation
conundrum} can be solved only if a fraction of the first generation of
massive stars have masses $\lsim 260 \msun$. Under such circumstances,
these will explode as \sngg and enrich the gas with heavy elements up
to a mean metallicity of $Z \gsim 10^{-4} Z_{\odot}$, and as per
arguments outlined in the previous section (see Fig.~1), thereafter
causing a shift over to an IMF that is similar to the local
one.  In what follows, we further explore the implications of the
above scenario and derive the conditions for the solution of the 
conundrum.

\subsection{Abundance of VMBHs and metals}

As a consequence of the picture proposed above, VMBHs are an
inevitable outcome. We now compute the expected mass density of metals
and the mass density of remnant VMBHs produced in such a first episode
of star formation. For a collapsed dark matter halo of total mass $M$,
the associated baryonic mass is assumed to be $M (\Omega_B/\Omega_M)$
(where $\Omega_B/\Omega_M$ is simply the baryon fraction). Following
the results of numerical simulations at different resolutions (Bromm \etal 2001, Abel \etal 2000) 
we assume that $\approx 1/2$ of the total
available gas is utilized in star formation, the rest remaining in
diffuse form.

The relative mass fraction of \sngg  and VMBH progenitors are
parametrized as follows:
\begin{eqnarray}
M_{\gamma\gamma} & = & f_{\gamma\gamma} \frac{M}{2}\,(\frac{\Omega_B}{\Omega_M}) = f_{\gamma\gamma} M_*,\\ \nonumber
M_{\bullet}& = & (1-f_{\gamma\gamma}) \frac{M}{2}\,(\frac{\Omega_B}{\Omega_M}) = (1-f_{\gamma\gamma}) M_*,
\end{eqnarray}
where $M_{\gamma\gamma}$ ($M_{\bullet}$) is the total mass which ends
up in \sngg (VMBHs) and $M_*$ is the mass processed into stars. Thus,
only a fraction $f_{\gamma\gamma}$ of the formed stars can contribute to gas metal
enrichment. The metal yields for the dominant elements have been
computed using the results of Heger \& Woosley (2001), and are plotted
in Fig.~3 as a function of the mass of the progenitor star. The bulk
of the yield is contributed by O$^{16}$ and the mass of metals ejected
can be written as \be M_Z \approx M_{\gamma\gamma}/2 - 10 \msun \frac{M_{\gamma\gamma}}{200 \msun},  \ee
where we have taken $\approx 200 \msun$ as a fiducial mass for \sngg progenitors.
Next, the further assumption is made that metals are ejected from the
parent galaxy into the IGM and their cosmic volume filling factor is
close to unity, therefore uniformly polluting the IGM.  This comes
from the results of Madau, Ferrara \& Rees (MFR, 2001) and Mori, Ferrara \&
Madau (2001), where it is shown that for reasonable values of the star
formation efficiency, metal bubbles produced by (standard SNII in)
proto-galaxies which result from 2-$\sigma$ peaks of the density power
spectrum at redshift 10 do overlap.  Indeed, the kinetic energy
released by \sngg is much higher than for ordinary SNII (see Fig.~2),
hence making our assumption even more solid. The temperature of the hot
gas is expected to be somewhat higher than estimated by MFR: however, due to
the higher redshifts and consequent stronger Inverse Compton cooling, ejected metals have enough
time to cool by $z \approx 3$.

As long as the average metallicity is below the critical value, \ie
$Z_{\rm cr} = 10^{-4} Z_{\odot}$, we argue that the IMF remains
top-heavy and the redshift-dependent critical density of metals
contributed by all \sngg at redshifts $>z$ can be computed, 
\be \Omega_Z (z) =
\frac{1}{\rho_{\rm cr}} \int_z^{\infty}dz'\int_{M_{min}(z')}
\\ dM\,n(M,z')\,M_Z 
\ee 
where $n(M,z)$ is the number density of halos per unit mass predicted
by the Press-Schechter formalism and the integration is performed from
$M_{\rm min}(z)$ which is the minimum mass that can cool within a
Hubble time at the specified formation redshift $z$, \ie 
$t_{\rm cool}(M,z) \lsim t_{H}(z)$ (see Ciardi \etal 2000). We adopt a
cosmological model with the following parameters to compute the
abundance of halos : $\Omega_{M} = 0.3$, $\Omega_{\Lambda} = 0.7$, $h
= 0.65$, $\Omega_B=0.047$ (latest predictions from BBN, see Burles,
Nollett \& Turner 2001) and use the COBE-normalized power spectrum for
fluctuations as described in Efstathiou, Bond \& White (1992).

From the above expression, we can estimate the transition redshift
$z_f$ at which the mean metallicity is $10^{-4} Z_{\odot}$, 
\be
<Z(z_f)> = \frac{\Omega_Z(z_f)}{\Omega_B} \approx 10^{-4}
Z_{\odot} 
\ee 
for various values of the fraction $f_{\gamma\gamma}$. 
The results are plotted in
Fig.~4 alongwith the corresponding
critical density $\Omega_{VMBH}$ contributed by the VMBHs formed,
which is given by, 
\be 
\Omega_{VMBH}(z_f) = \frac{(1 - f_{\gamma\gamma})}{\rho_{\rm
crit}} \int_{z_f}^{\infty} dz'\int_{M_{min}(z')}
dM\,n(M,z')\,M_*.  
\ee 
Several assumptions as to the fate of these
VMBHs at late times will be examined in an attempt to obtain
constraints on the value of $f_{\gamma\gamma}$ 
from current observations in the next Section.

\section{Observational Constraints}
 
The objective now is to set some limits on $f_{\gamma\gamma}$, the fraction
of stars formed that result in \sngg providing the first metals, that
are eventually responsible for the shift to a normal IMF.  In order to
compare the predicted critical density of VMBH remnants to present
observational data we need to make further assumptions about the fate
of these VMBHs at late times.

The mass density contributed by remnant supermassive black holes
$\Omega_{SMBH}^0$ can be estimated observationally\footnote{The
superscript $0$ denotes the present day value} from the demography of
local galaxies (Magorrian et al. 1998; Merritt \& Ferrarese 2001).
There are two extreme possibilities (bounding cases for the values of
$f_{\gamma\gamma}$) concerning the relation between the inferred mass
density of SMBHs and VMBHs:

[A] $\Omega_{VMBH}(z_f) = \Omega_{SMBH}^0$:
all VMBHs have, during the course of galaxy mergers, in fact,
been used to build up the SMBHs detected today; 

[B] $\Omega_{VMBH}(z_f) = \Omega_{VMBH}^0  \ne \Omega_{SMBH}^0$
there is no relation between the detected SMBHs and the early
VMBHs, implying that SMBHs have grown primarily by gas accretion.

Thus, in the first scenario [A], we assume that all VMBHs are able to
sink to the center of host systems to merge within a Hubble time,
giving birth to the SMBHs in the nuclei of galaxies that we see
today. Note, that this is in fact unlikely (as explained further
below), nevertheless, this argument provides an extreme bound. 

It is important to point out here, that in the current proposed
theoretical models for the mass accretion history of black holes, the
mass density seen today locally in SMBHs can be built either
predominantly by mergers or by accretion. In phenomenological models
that attempt to tie in the current high redshift observations of the
space density and luminosity function of quasars with that of the
local space density of black holes wherein it is assumed that quasars
are {\it (i)} powered by BHs and {\it (ii)} are optically bright for a
period of $\approx 10^{6-7}$ yr (Haiman \& Loeb 1998; Haehnelt,
Natarajan \& Rees 1998; Kauffmann \& Haehnelt 2000), either picture
(mergers/accretion) can adequately explain the mass assembly of the
remnant SMBHs.

This permits both scenarios [A] (wherein SMBHs grow primarily via
mergers of VMBHs during galaxy assembly) and [B] (wherein SMBHs grow
primarily by accretion and $\Omega_{SMBH}^{0}$ is unrelated to
$\Omega_{VMBH}$) to be plausible extreme cases.  Consideration of
cases [A] and [B] will provide bounds on the value for $f_{\gamma\gamma}$.
For a given $f_{\gamma\gamma}$, the corresponding value for the 
transition redshift can be found on the $z_f$ curve shown in Fig~4.

We note here that VMBHs in the mass
range expected from the first episode of metal-free star formation (a
few hundred solar masses) are likely to be ubiquitous. Using ISO
observations, Gilmore \& Unavane (1998) have convincingly argued that
low mass stars (in fact hydrogen burning stars of any mass) do not
provide a substantial portion of the dark matter inferred in the halos
of galaxies. In combination with the mass limits obtained on potential
dark matter candidates from the MACHO (Alcock \etal 2001) and EROS 
(Lasserre \etal 2000) experiments (micro-lensing studies) in our Galaxy, 
low mass objects, with masses under $10\,M_{\odot}$ are ruled out 
- thereby, making VMBHs ($\approx 100
\msun$ BHs) in halos (case [B]) plausible dark matter candidates.

For case [A], we can simply equate $\Omega_{VMBH}^0$ to the measured
mass density in SMBHs found from the demography of nuclei of nearby
galaxies (Magorrian \etal 1998; Gebhardt \etal 2001) indicated by
the horizontal line in Fig.~4 (top panel). 
The most recent value reported by Merritt \& Ferrarese (2001) 
is $ \Omega_{SMBH}^0 = 10^{-4} h
\Omega_B$, which for $h = 0.65$ and $\Omega_B= 0.047$ yields
$\Omega_{SMBH}^0 = 3.05 \times 10^{-6}$. From the top panel of Fig.~4, 
we see that this
gives $f_{\gamma\gamma} \approx 0.06$.
The corresponding value for the transition redshift can be found 
on the $z_f$ curve on the same plot, yielding a value 
$z_f \approx 18.5$. This constraint implies that all but
a small fraction of the mass involved in the first episodes of star
formation - approximately 6\% - went in objects outside the 
mass range $140 \msun < M_{\star} < 260 M_{\odot}$, yielding remnant black
holes with mass $10 - 500\,M_{\odot}$. 
The scenario corresponding to [A] therefore implies that $z_f$, the
transition redshift from top-heavy to normal IMF occurred at $\approx$
18.5.

Let us now look at the other extreme case [B], wherein the assembled
SMBHs in galactic centers have formed primarily via accretion and are
unrelated to VMBHs. This leads us to two further possibilities: (i)
VMBHs would still be in the process of spiralling into the centers of
galaxies due to dynamical friction but are unlikely to have reached
the centers within a Hubble time due to the dynamical friction
time-scale being long (Madau \& Rees 2001), or (ii) VMBHs contribute
the entire baryonic dark matter in galactic halos.

Pursuing scenario (i), therefore, at best some fraction of these VMBHs
(for instance, those in gas-rich regions of gas-rich systems) might
appear as off-center accreting sources that show up in the hard X-ray
wave-band. Such sources have indeed been detected both by ROSAT
(Roberts \& Warwick 2000) and more recently by Chandra in nearby
galaxies (M84: Jones 2001; NGC 720: Buote 2001).

In a survey of archival ROSAT HRI data to study the X-ray properties
of the nuclei of 486 optically selected bright nearby galaxies (Ho,
Fillipenko \& Sargent 1995), Roberts \& Warwick (2000) found a large
number of off-center X-ray sources. The X-ray sources detected within
the optical extent of these galaxies were classified either as nuclear
or non-nuclear (and therefore off-center) depending on whether the
source was positioned within 25 arcsec of the optical nucleus. They
detect a nuclear source in over 70\% of the galaxy sample and a total
of a 142 off-center sources.  Roberts \& Warwick (2000) find that the
non-nuclear sources follow a steep, near power-law X-ray luminosity
distribution in the $10^{36} - 10^{40}\mbox{~erg s}^{-1}$ which leads
to an $L_X/L_B$ ratio of \be \frac{L_X}{L_B} = 1.1 \times
10^{39}\mbox{~erg s}^{-1} (10^{10} \mbox{~L}_{\odot})^{-1}.
\label{eq:averageLx}
\ee 

The median luminosity of the non-nuclear sources is found to be a
factor of $\sim 10$ lower than that of the nuclear sources. However,
they estimate the incidence rate of off-center sources with $L_X \ge
10^{38.3}\mbox{~erg s}^{-1}$ (which corresponds to the Eddington
luminosity for a 1.4 solar mass neutron star) to be $\approx 0.7\,{\rm
per}\,10^{10}\,L_{\odot}$ galaxy. The existence of accreting VMBHs
might also help explain the following observation. The far-IR excess
detected with DIRBE at 60 and 100 $\mu$m has been tentatively
interpreted as an extra-galactic background with integrated intensity
of $44 \pm 9$ nW $\mbox{m}^{-2} \mbox{sr}^{-1}$ in the range $60-100
\mu$m (Finkbeiner, Davis \& Schlegel 2000). The energy required to
produce such a FIR background could derive from a highly obscured
population of accreting VMBHs at moderate redshifts (for an
alternative to this see explanation of scenario (ii) below).

In order to estimate the mass density contributed by these VMBHs, we
can compare the mass in VMBHs to that of the SMBH in a typical galaxy
of luminosity say $10^{10}\,L_{\odot}$ for case [B]. Most of the VMBHs
are likely to be en-route to galactic centers.  Note however, that not
all the inspiralling VMBHs will be accreting and be X-ray bright: many
of these could, in fact, be low radiative efficiency ADAFs (Advection
Dominated Accretion Flows), and therefore be too faint to be detected
by ROSAT or Chandra or not accreting at all if they do not happen to
be in gas-rich regions of the galaxy. Thus, using the following 
argument we can obtain a lower limit on the abundance of VMBHs.

The mass of the central SMBH hosted by a galaxy with a luminosity
$L_{galaxy} = 10^{10}\,L_{\odot}$ (assumed to be the fiducial galaxy
for these purposes) in the B-band is given by (Merritt \& Ferrarese
2001):
$$M_{SMBH} =  10^{-3} M_{\rm bulge} =    10^{-3}
L_{\rm galaxy}$$ 
where the mass-to-light ratio in the B-band is taken to be $1\,
M_{\odot}/L_{\odot}$ and the galaxy luminosity is essentially
dominated by the bulge luminosity. Hence, the mass of the central SMBH
is $M_{SMBH} \approx 10^7 M_{\odot}$.  Now, we use the luminosity of
the ROSAT off-center sources (Roberts \& Warwick 2000) and use the
fact that they find $\approx$ 0.7 off-center sources with luminosity
$\ge 10^{38.3}\mbox{~erg s}^{-1}$ per $10^{10}\,L_{\odot}$ galaxy, to
estimate the mass of VMBHs in the fiducial galaxy. If we assume that
these VMBHs have a typical mass of $\approx 300 \msun$, their
Eddington luminosity is,
$$L_{\rm Edd} = 4.3 \times 10^{40} (\frac{M_{VMBH}}{300 \msun})  
\mbox{~erg s}^{-1}.$$ 
Thus, VMBHs appear to be radiating at sub-Eddington
luminosities, the average rate (given by eq.~\ref{eq:averageLx}) 
being $\approx 3 \%$ of the Eddington value. This is consistent
with the observed luminosities of the nuclear sources in the sample,
which Roberts \& Warwick (2000) find to be radiating
at severely sub-Eddington rates.

Therefore, we expect that the typical accreting VMBH mass in a
$10^{10}\,L_{\odot}$ galaxy to be about 210$\,M_{\odot}$.  Little is
known about the spatial distribution of such objects.  How can we take
into account the fact that only a fraction of the sources are likely
to be accreting? Since the detected ROSAT off-center sources are
within the optical radius of the galaxies, the number of non-accreting
VMBHs in the halos of these galaxies can be large.  One can obtain a
simple estimate of this number by pursuing the following
argument. Assume that VMBHs, being collisionless particles, closely
trace the dark matter (which we take to obey a Navarro, Frenk \& White
[NFW, 1997] profile) and that the ratio of virial to optical
radius is $\approx 15$ for a $10^{10}\,L_{\odot}$ disk 
galaxy (Persic, Salucci \& Stel,
1996).  These hypotheses require that we scale up the total mass in
VMBHs by a factor of $\approx 10$, which gives $M_{VMBH} = 2.1
\times 10^3\,M_{\odot}$. Now, the ratio $M_{VMBH}/M_{SMBH}$ is
$\approx 2.1 \times 10^{-4}$, implying that $\Omega_{VMBH} = 2.1
\times 10^{-4} \Omega_{SMBH}^0 = 6.4 \times 10^{-10}$ which in turn
gives $f_{\gamma\gamma} \approx 1$ (see
dashed line in bottom panel of Fig.~4) and $z_f \sim 22.1$.

Now we explore scenario (ii) of Case [B], the instance when VMBHs
constitute the entire baryonic dark matter content of galaxy halos
(but do not contribute to the disk dark matter). It is important 
to point out here that cosmological nucleosynthesis arguments
require both baryonic and non-baryonic dark matter (Pagel 1990).
Essentially, this is due to the fact that the mass density contributed 
by baryons $\Omega_B$, is well in excess of $\Omega_V$, the contribution 
to the mass density by visible baryons.  

Using the luminosity dependent relation of visible to dark matter 
for spirals (Persic, Salucci \& Stel, 1996) for a fiducial galaxy 
of $10^{10} \Ls$ we find,
\[
\frac{M_{vis}}{M_{DM}} \approx 0.05
\]
where $M_{vis}$ is the visible mass and $M_{DM}=M_{DM}^b+M_{DM}^{nb}$
is the total dark matter mass, given as a sum of a baryonic and 
a non-baryonic component. The total baryonic mass is a fraction 
$\Omega_B/\Omega_M$ of the total mass of the system,
\eg
\[
M_{vis}+M_{DM}^{b}=\frac{\Omega_B}{\Omega_M} (M_{vis}+M_{DM}).
\]
From the two equations above and assuming a NFW density profile,
we estimate that the total baryonic dark matter content as,
\[
M_{DM}^b \simeq (\Omega_M^{-1}-1) M_{vis}  = 2.33 M_{vis};
\]   
moreover, 90\% of this mass resides outside the optical radius and can
be contributed by VMBHs:
\be 
\frac{M_{VMBH}}{M_{SMBH}} = \frac{2.1 \times M_{vis}}{10^{-3} \times
M_{vis}}= 2.1 \times 10^3,\,\,\,
\ee
implying
\be
\Omega_{VMBH} = 2.1\times 10^3  \times \Omega_{SMBH} = 6.4 \times
10^{-3}.  
\ee 
Incorporating this constraint into Fig. 4 (bottom panel),
we find  $f_{\gamma \gamma} \approx 3.15 \times 10^{-5}$ and $z_f \sim
5.4$. Note that this scenario does not violate the constraint on the
over-production of background light (see Carr 1998; Bond, Carr \&
Hogan 1991; Wright et al. 1994). In fact, some fraction of the
observed near-IR DIRBE excess could be produced by VMBHs. 
In a recent estimate of the cosmic background at 1.25 $\mu$m and
2.2 $\mu$m (corresponding to the J and K bands respectively) using
2MASS and the DIRBE results Cambresy et al. (2001), also find an
excess (significantly higher than the integrated galaxy counts in the J and
K bands) suggesting the contribution of other sources. Pop III stars
and their VMBH remnants (accreting at very high redshifts) postulated 
here are likely candidates.  

According to scenario [B], the average metal abundance at redshift
$\sim 5.4$ should be $Z_{\rm cr} = 10^{-4} Z_{\odot}$, marking the
transition from a top-heavy IMF to a standard power-law IMF. This does
not necessarily violate the observed metal abundances in Damped
Ly$\alpha$ systems, $Z \approx 10^{-3}$, as it is yet not clear to
what extent the IGM metallicity is spatially uniform at these
intermediate redshifts.

Therefore, scenario [B] is poorly constrained by the data, as the
two options (i) and (ii) give large ranges for 
$f_{\gamma\gamma}$ and $z_f$: 
$3.15 \times 10^{-5} <f_{\gamma\gamma} < 1$ and $5.4< z_f < 22.1$.
Consistently, scenario [A] gives a limit on $f_{\gamma\gamma}$ and $z_f$ of 
$f_{\gamma\gamma} = 0.06$ and $z_f = 18.5$.

The actual data do not
allow us at present to strongly constrain the above two quantities,
but they provide interesting bounds on the proposed scenario.  These
ranges also have implications for the expected detection rate of SNe
beyond $z_f$ with future instruments like NGST (Marri, Ferrara \& 
Pozzetti 2000).

\section{Discussion}

We have proposed a scenario to solve a puzzling {\it star formation
conundrum}: the first stars are now thought to be very massive and
hence to lock their nucleosynthesis products into a remnant (very
massive) black hole.  This high-mass biased star formation mode
continues as long as the gas remains metal-free. During this phase,
metal enrichment can occur only if a fraction $\fgg$ of the stars have
mass in the window leading to pair-unstable SNe ($140 \msun < M_\star
< 260 \msun$) which disperse their heavy elements into the surrounding
gas.  Such metals enrich the gas up to $Z_{\rm cr}\approx 10^{-4}\,Z_{\odot}$,
when a transition to efficient cooling-driven fragmentation producing
$ \simlt 1 \msun$ clumps occurs at redshift $z_f$.  We argue that the
remaining fraction of the first stars end up in $\approx 100 \msun$
VMBHs.  By analyzing the evolutionary fate of such objects, we argue
that [A] they could end up in the SMBHs in the centers of galactic
nuclei, [B] (i) could be en-route to the center and hence identified with
the X-ray bright, off-center ROSAT sources, or (ii) constitute the
entire baryonic dark matter content of galaxy halos.  These
possibilities are used to obtain constraints on the two quantities:
$f_{\gamma\gamma} \sim 0.06$ and $z_f \sim 18.5$ for
[A], and $f_{\gamma\gamma} \approx [10^{-5}-1]$ and $z_f \approx [5.4-22.1]$ 
for case [B].  
The value $Z_{\rm cr} \approx 10^{-4} Z_\odot$ found here is admittedly 
somewhat uncertain. For this reason we have investigated how the above results
might be affected by a different choice, \eg assuming $Z_{\rm cr} \approx 10^{-5}
Z_\odot$. Indeed, lower values of  $Z_{\rm cr}$ imply that less efficient metal enrichment
is required in order to change to a more conventional star formation mode. Thus,
comparable values for $z_f$ are found but the corresponding values for $\fgg$ 
tend to be systematically smaller, being $\fgg \approx 0.6\%$ for scenario [A] and in the range
$3.15 \times 10^{-6}<z_f<0.98$ for scenario [B].   

Several uncertainties remain in the comparison of the inferred density
of VMBHs to local observations.  For example, it is not obvious if
SMBHs could at all form out of VMBHs. Dynamical friction can
effectively drag VMBHs towards the center of the host system, at least
within a distance of $\sim 100$~pc (Madau \& Rees 2001).  Unless most
of the energy is radiated away in gravitational waves, it could be
difficult for a VMBH cluster to coalesce into a single unit.  
 Furthermore, accretion onto isolated VMBHs could be too
inefficient to explain most of the off-center sources observed by
ROSAT and CHANDRA. Higher accretion rates might be activated by the
tidal capture/disruption of ordinary stars. Question remains if the
frequency of such an event is in fact sufficiently high to explain the data.

In spite of the many difficulties and uncertainties discussed above,
our study represents a first attempt to link the first episode of
cosmic star formation activity to present day observational evidence
of their fossil remnants.

A top-heavy IMF for the early episodes of star formation in the
universe might have other interesting observational consequences, we
speculate further on them below.  The kinetic energy released during
the thermonuclear explosions powered by pair instability are $\approx
10^2$ larger than those of ordinary Type II SNe. This might cause the
interaction with the circumstellar medium to be as strong as predicted
for hypernovae (Woosley \& Weaver 1982).  However, these explosions
do not lead to the ejection of strongly relativistic matter (Fryer,
Woosley \& Heger 2001) and therefore cannot power a gamma-ray burst
(GRB).
 
In PopIII progenitors of VMBHs, the estimated angular momentum is
sufficient to delay black hole formation and the system might develop
triaxial deformations (Fryer, Woosley \& Heger 2001). If the
instabilities have enough time to grow, the core might break into
smaller fragments that would then collapse and merge to form the
central VMBH. If not, the star might still develop a bar-like
configuration. Both these scenarios lead to a significant emission of
gravitational waves (Schneider \etal 2000; Fryer, Holz \& Hughes
2001). Furthermore, significant emission of gravitational radiation
can occur as a result of the in-spiral and merger of VMBHs onto the SMBHs
in the center of host systems (Madau \& Rees, 2001)
 
Once the VMBHs have formed, accretion continues through a disk at a
rate which can be as large as $10 \msun \mbox{s}^{-1}$ (Fryer, Woosley
\& Heger 2001).  Magnetic fields might drive an energetic jet which
can produce a strong gamma-ray burst through the interaction with
surrounding gas.  The properties of these PopIII GRBs would be
considerably different from their more recent ($z<5$) counterparts:
depending on the uncertain interaction of the jet with the surrounding
matter, the bursts would be probably longer [$10 (1+z)$~s] and the
peak of emission, that in the rest frame is in $\gamma$-rays, would be
shifted into X-rays.  Indeed, BeppoSAX has revealed the existence of a
new class of events, the so-called X-ray flashes or X-ray rich GRBs,
which emit the bulk of their energy in X-rays (Piro, private
communication).  Furthermore, since PopIII GRBs explode at very high redshifts,
it is likely that their optical afterglow might be heavily absorbed by
the intervening gas. These systems might be the natural candidates for
the significant number (about 40 \% of GRBs for which fast follow-up
observations were carried out) of GRBs which do not show an optical
counterpart, the so-called GHOST (GRB Hiding Optical Source Transient)
or dark GRBs. Other explanations ascribe the failed optical detection
to dust extinction within the host system but the ultimate nature of
GHOSTS is still very debated (Lazzati, Covino \& Ghisellini, 2001;
Djorgovski \etal 2001)
 
Finally, the energetic jets generated by GRBs engines produce,
by photon-meson interaction, a burst of TeV neutrinos while propagating
in the stellar envelope (M{\'e}sz{\'a}ros \& Waxman 2001). We investigate 
this aspect in a companion paper (Schneider, Guetta \& Ferrara 2002) where we
use the constraints set by the AMANDA-B10 experiment on the total 
integrated flux of TeV neutrinos from PopIII GRBs.

\begin{acknowledgments}
We are grateful to P. Coppi, K. Nomoto \& M. Rees for comments and 
suggestions. We wish to thank V. Bromm  and L. Piro for discussions.
We thank the referee for his/her thoughtful comments. 
This work is partially supported (RS) by the Italian CNAA (Project 16/A);   
KO is supported by Research Fellowships of the Japan
Society for the Promotion of Science for Young Scientists, grant 6819.

\end{acknowledgments}

\begin{figure}
\centerline{\psfig{figure=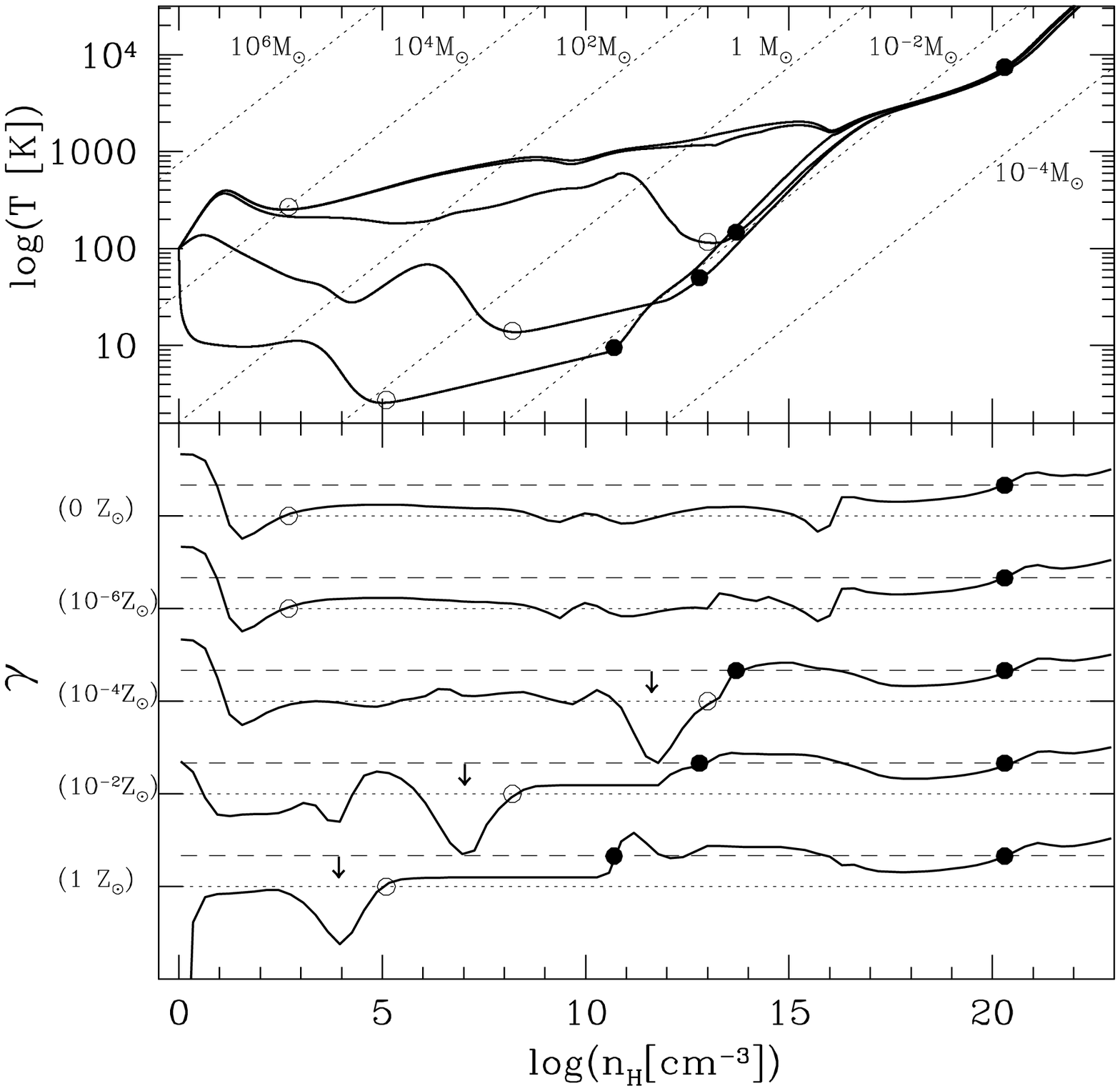,height=17cm}}
\caption{\footnotesize {\it Upper panel}: The evolution of the
temperature as a function of the hydrogen number density of
proto-stellar clouds with the same initial gas temperature but
varying metallicities $Z = (0, 10^{-6}, 10^{-4}, 10^{-2}, 1) Z_\odot$
($Z$ increasing from top to bottom curves). The dashed lines correspond to
the constant Jeans mass for spherical clumps; open circles indicate the
points where fragmentation stops, filled circles mark the formation of
hydrostatic cores. This Figure is reproduced from Figure 1 of Omukai (2000) for illustration after
some modifications. {\it Lower panel}: The adiabatic index $\gamma$ as a
function of the hydrogen number density for the curves shown in the upper
panel.  Dotted (dashed) lines correspond to $\gamma = 1$
($\gamma=4/3$); open and filled circles as above.}
\end{figure}

\begin{figure}
\centerline{\psfig{figure=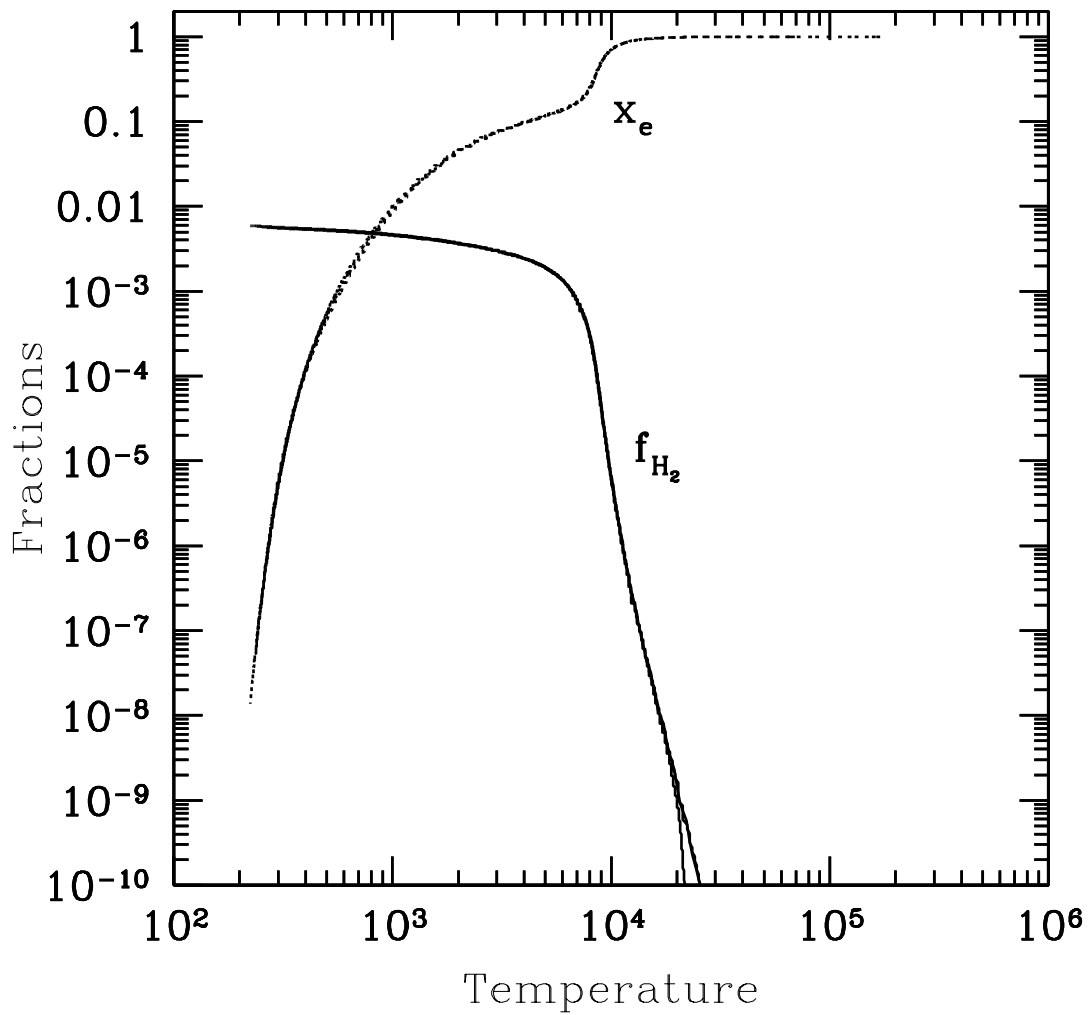}}
\caption{\footnotesize Fractional abundance of H$_2$ (solid lines) 
and ionization fractions
(dashed lines) as a function of gas temperature. The three cases plotted are 
for a halo of mass $M=10^8 M_\odot$ at three different virialization redshifts 
$z = 15, 20, 25$, (virial temperatures of $T_{\rm vir} = [2.2, 6.9, 17] 
\times 10^4$~K). As can be seen in the Figure, the evolution appears to be  
independent of the virial temperature of the halo.}
\end{figure}
\begin{figure}
\centerline{\psfig{figure=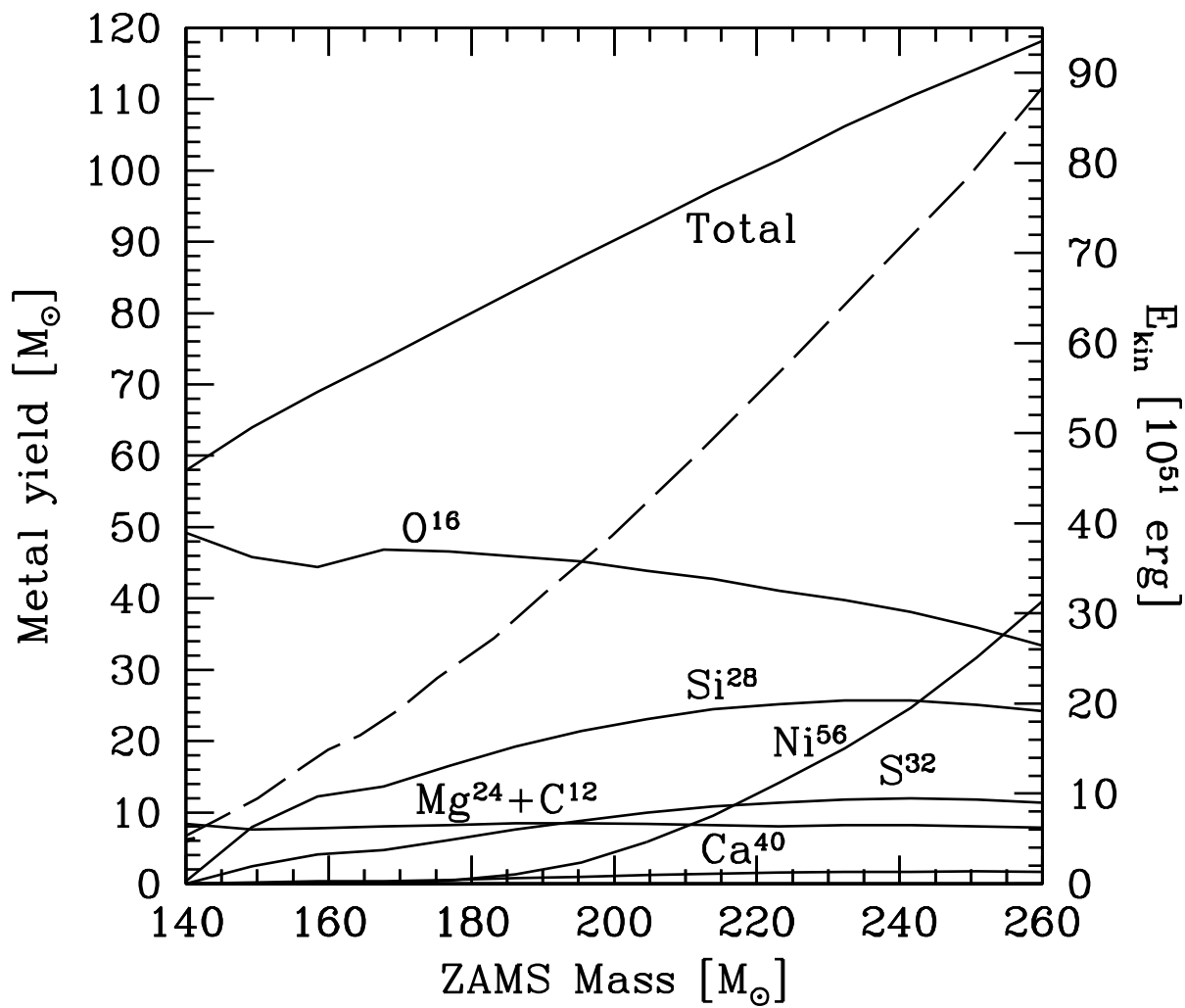}}
\caption{\footnotesize Metal yields of the main elements 
synthesized in metal-free \sngg              
according to Heger \& Woosley (2001). The upper solid line corresponds to the
total yield ejected and the dashed line indicates the kinetic energy of the explosion.}
\end{figure}

\begin{figure}
\centerline{\psfig{figure=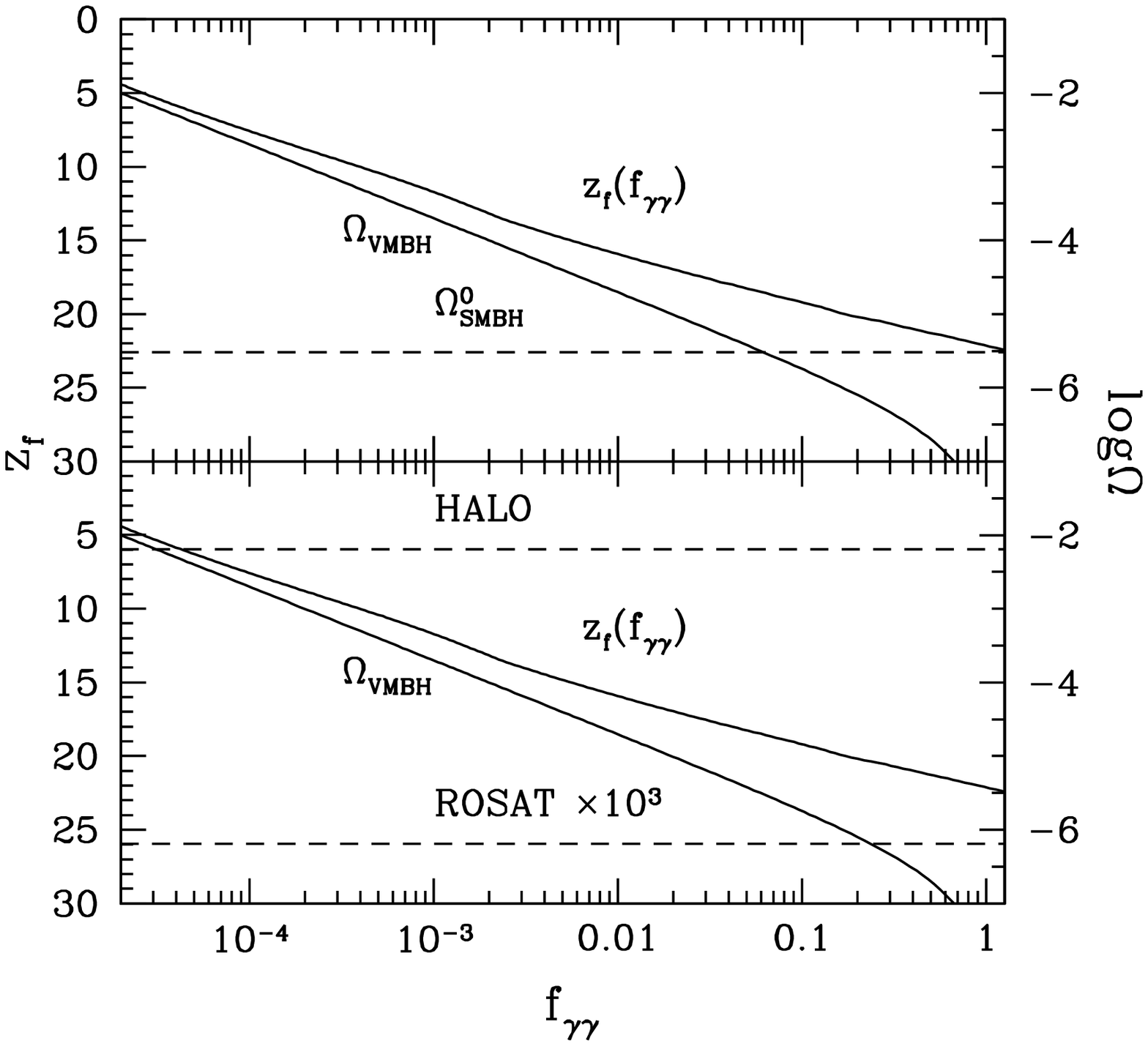,width=15.cm}}
\caption{\footnotesize
The top-heavy to normal IMF transition redshift, $z_f$,
as a function of \sngg progenitor mass fraction 
and the mass density contributed by VMBHs $\Omega_{VMBH}$. 
{\it Upper Panel}: The computed critical density of VMBH remnants 
is compared to the observed value for SMBHs (upper dashed horizontal line).
{\it Lower Panel}: The computed critical density of VMBH remnants 
is compared to the contribution to $\Omega$ from the X-ray bright, off-center 
ROSAT sources (the lower dashed line) and to the abundance predicted assuming
that the baryonic dark matter in galaxy halos is entirely contributed by VMBHs 
(upper dashed line). The observations on $\Omega_{VMBH}$ 
constrain the value for $f_{\gamma\gamma}$. For a given $f_{\gamma\gamma}$, 
the corresponding value for the transition redshift
can be inferred by the $z_f$ curve.} 
\end{figure}


\begin{references}

\reference{}
Abel T., Anninos P., Norman M. L., \&  Zhang Y. 1998, ApJ, 518

\reference{} 
Abel, T., Bryan, G. \& Norman, M. 2000, ApJ, 540, 39

\reference{}
Alcock, C. \etal 2001, ApJ, 550, L169

\reference{}
Bond, J. R., Carr, B. J., \& Hogan, C., 1991, ApJ, 367, 420

\reference{}
Bromm V., Coppi P.S., \& Larson R.B., 1999, ApJ, 527, L5

\reference{}
Bromm V., Ferrara A., Coppi P.S., \& Larson R.B., 2001, MNRAS, 328, 969

\reference{}
Buote, D., 2001, Proceedings of the Yale Workshop on galaxy shapes, ed. Natarajan, 
World-Scientific Publishers, New York

\reference{}
Burles, S., Nollett, K., \& Turner, M. S., 2001, ApJ, 552, L1

\reference{}
Cambresy, L., Reach, W. T., Beichmann, C. A., \& Jarrett, T. H., 2001,
ApJ, 555, 563

\reference{}
Carr, B. J., Bond, J. R., \& Arnett, W. D., 1984, ApJ, 277, 445

\reference{}
Carr, B. J., 1998, Phys. Rep., 307, 83

\reference{}
Ciardi, B., Ferrara, A. \& Abel, T.  2000, ApJ, 533, 594

\reference{}
Ciardi, B., Ferrara, A., Governato, F., Jenkins, A., 2000, MNRAS, 314, 611

\reference{}
Djorgovski, S. G., Frail, D. A., Kulkarni S. R., Bloom J. S., Odewahn S. C.,
 Diercks A., 2001, ApJ, submitted, (astro-ph/0107539)


\reference{}
Efstathiou, G. P., Bond, J. R., \& White, S. D. M., 1992, MNRAS, 258, 1

\reference{}
El Eid, M. F., Fricke, K. J. \& Ober, W. W. 1983, A\&A, 119, 54

\reference{}
Finkbeiner, D., Davis, M., \& Schlegel, D. J., 2000, ApJ, 544, 81

\reference{}
Fowler, W. A. \& Hoyle, F. 1964, ApJS, 9, 201

\reference{}
Fricke, K. J. 1973, ApJ, 183, 941

\reference{}
Fryer, C. L., 1999, ApJ, 522, 413

\reference{}
Fryer, C. L., Woosley, S. E. \& Heger, A., 2001, ApJ, 550, 372

\reference{}
Fryer, C. L., Holz D. E. \& Hughes S. A., 2001, ApJ, submitted, (astro-ph/0106113)

\reference{}
Fuller, G. M., Woosley, S. E. \& Weaver, T. A. 1986, ApJ, 307, 675


\reference{}
Gebhardt, K., Kormendy, J., Ho, L.~C., et al., 2001, ApJ, 543, L5

\reference{}
Gilmore, G., \& Unavane, M., 1998, MNRAS, 301, 813

\reference{}
Haehnelt, M., Natarajan, P., \& Rees, M. J., 1998, MNRAS, 300, 817

\reference{}
Haiman, Z., \& Loeb, A., 1998, ApJ, 503, 505

\reference{}
Haiman, Z., Thoul, A., \& Loeb, A., 1996, ApJ, 464, 523 

\reference{}
Heger, A. \& Woosley, S. E., 2001, ApJ, in press, (astro-ph/0107037)

\reference{}
Hernandez, X. \& Ferrara, A., 2001, MNRAS, 324, 484

\reference{}
Ho, L., Fillipenko, A., \& Sargent, W., 1995, ApJS, 98, 477

\reference{}
Jones, C., 2001, Proceedings of the Yale Workshop on galaxy shapes, ed. Natarajan, 
World-Scientific Publishers, New York

\reference{}
Kauffmann, G., \& Haehnelt, M., 2000, MNRAS, 318, L35

\reference{}
Kroupa, P. 2001, MNRAS, 322, 231

\reference{}
Larson, R. B., 1999, Proc. of Star Formation 1999, Ed. Nakamoto, 
Nobeyama Radio Observatory, 336

\reference{}
Lasserre, T., \etal 2000, A \& A, 355, L39

\reference{}
Lazzati, D., Covino, S. \& Ghisellini, G., 2002, MNRAS, 330, L583

\reference{}
Loewenstein, M., 2001, ApJ, 557, L35

\reference{}
Madau, P., Ferrara, A. \& Rees, M. J. 2001, ApJ, 555, 92

\reference{}
Madau, P. \&  Rees, M. J. 2001, ApJL, 551, 27 

\reference{}
Magorrian, J., et al., 1998, AJ, 115, 2285  

\reference{}
Marri, S.,  Ferrara, A. \& Pozzetti, L. 2000, MNRAS,
317, 265

\reference{}
Merritt, D. \& Ferrarese, L., 2001, MNRAS, 320, L30

\reference{}
M{\'e}sz{\'a}ros, P.\& Waxman, E.,  2001, Phys. Rev. Lett., 87, 171102 

\reference{}
Mori, M., Ferrara, A., \& Madau, P. 2001, ApJ, submitted
(astro-ph/0106107)

\reference{} 
Nakamura, F. \& Umemura, M. 1999, ApJ, 515, 239

\reference{} 
Nakamura, F. \& Umemura, M. 2001, ApJ, 548, 19

\reference{} 
Navarro, J., Frenk, C. S., \& White, S. D. M., 1997, ApJ, 490, 493

\reference{}
Norris, J. E., Beers, T. C., \& Ryan, S. G., 2000, ApJ, 540, 456

\reference{}
Oh, S.P., Haiman, Z., 2001, ApJ, submitted, (astro-ph/0108071)

\reference{} 
Omukai, K. \& Nishi, R. 1998, ApJ, 508, 141

\reference{}
Omukai, K. 2000, ApJ, 534, 809 

\reference{}
Omukai, K. 2001, ApJ, 546, 635

\reference{}
Omukai, K. \& Inutsuka, S. 2001, MNRAS, submitted, (astro-ph/0112345) 

\reference{}
Oh, S. P., Nollett, K. M., Madau, P. \& Wasserburg, G. J. 2001, ApJL,
submitted, (astro-ph/0109400)

\reference{}
Pagel, B. E., 1990, Phy. Rep., 333, 433

\reference{}
Persic, M., Salucci, P. \& Stel, F., 1996, MNRAS, 281, 27

\reference{} 
Rees, M. 1976, MNRAS, 176, 483

\reference{}
Rees, M.~J., \& Ostriker, J.~P., 1977, MNRAS, 179, 541

\reference{}
Roberts, T. P., \& Warwick, R. S., 2000, MNRAS, 315, 98

\reference{}
Ripamonti, E.,  Haardt, F., Ferrara, A. \& Colpi, M. 2001,
MNRAS, submitted, (astro-ph/0107095)

\reference{}
Scalo, J. 1998, The Stellar Initial Mass Function (38th Herstmonceux
Conference), eds. G. Gilmore \&  D. Howell, ASP Conf.
Series, 142, 201

\reference{}
Schneider, R., Ferrara, A., Ciardi, B., Ferrari, V., Matarrese, S., 2000,
 MNRAS, 317, 385

\reference{} 
Schneider, R., Guetta, D., Ferrara, A. 2002, submitted, astro-ph/0201342

\reference{}
Silk, J. 1977, ApJ, 211, 638

\reference{} 
Silk, J. 1983, MNRAS, 205, 705

\reference{}
Spitzer, L., 1978, Physical processes in the interstellar medium, New York 
Wiley-Interscience.

\reference{}
Stahler, S. W., Shu, F. H. \& Taam, R. E., 1980, ApJ, 241, 637

\reference{}
Susa, H., Uehara, H., Nishi, R. \& Yamada, M., 1998, Prog. Theor. Phys., 100, 63

\reference{}
Todini, P. \& Ferrara, A., 2001, MNRAS, 325, 726

\reference{}
Uehara, H., Susa, H., Nishi, R. \& Yamada, M., Nakamura, T. 1996, ApJ, 473, L95

\reference{}
Uehara, H., Inutsuka, S., 2000, ApJ, 513, L91


\reference{}
Umeda, H. \& Nomoto, K. 2002, to appear in ApJ (pre-print astro-ph/0103241)


\reference{}
Wright, E. L. \etal 1994, ApJ, 420, 450

\reference{} 
Wolfire, M. G., \& Cassinelli, J. P. 1987, ApJ, 319, 850

\reference{}
Woosley, S. E. \& Weaver, T. A., 1982, in Superovae: A Survey of Current
Research, ed. M. Rees \& R. J. Stoneham (Dordrecht: Reidel), 79

\reference{}
Woosley, S. E., \& Weaver, T. A., 1995, ApJS, 101, 181


\end{references}
\end{document}